\documentclass[useAMS,usenatbib]{mn2e}
\usepackage{epsfig}

\title[Pulsation models for the roAp star HD 134214]{Pulsation models for the roAp star HD 134214}
\author[H. Saio et al.]{H. Saio$^{1}$\thanks{E-mail:saio@astr.tohoku.ac.jp}, 
M. Gruberbauer$^{2}$\thanks{E-mail: mgruberbauer@ap.smu.ca}, 
W. W. Weiss$^{3}$\thanks{E-mail: werner.weiss@univie.ac.at}, 
J. M. Matthews$^{4}$\thanks{E-mail:matthews@astro.ubc.ca}, and
T. Ryabchikova$^{5}$\thanks{E-mail:ryabchik@inasan.ru}
\\
$^{1}$Astronomical Institute, Graduate School of Science, Tohoku University,
Sendai 980-8578, Japan\\
$^{2}$Department of Astronomy and Physics, Saint Mary's University, Halifax,
NS B3H 3C3, Canada\\
$^{3}$Institute for Astronomy, University of Vienna, T\"urkenschanzstrasse 17,
A-1180, Vienna, Austria\\
$^{4}$Department of Physics and Astronomy, University of British Columbia, \\
6224 Agricultural Road, Vancouver, BC V6T 1Z1, Canada\\
$^{5}$Institute of Astronomy, Russian Academy of Sciences, Pyatnitskaya 48, 119017 Moscow, Russia
}

\begin{document}

\date{Accepted  Received}

\pagerange{\pageref{firstpage}--\pageref{lastpage}} \pubyear{2011}

\maketitle

\label{firstpage}

\begin{abstract}
Precise time-series photometry with the MOST satellite has led to identification of 
10 pulsation frequencies in the rapidly oscillating Ap (roAp) star HD~134214. We 
have fitted the observed frequencies with theoretical frequencies of axisymmetric 
modes in a grid of stellar models with dipole magnetic fields.  We find that, among 
models with a standard composition of $(X,Z) = (0.70,0.02)$ and with suppressed 
convection, eigenfrequencies of a $1.65\,{\rm M}_\odot$ model with 
$\log T_{\rm eff} = 3.858$ and a polar magnetic field strength of 4.1kG agree 
best with the observed frequencies.  
We identify the observed pulsation frequency with the largest amplitude as 
a deformed dipole ($\ell = 1$) mode, and the four next-largest-amplitude 
frequencies as deformed $\ell = 2$ modes. 
These modes have a radial quasi-node in the outermost 
atmospheric layers ($\tau \sim 10^{-3}$). 
Although the model frequencies agree roughly 
with observed ones, they are all above the acoustic cut-off frequency for the model
atmosphere and hence are predicted to be damped. The excitation mechanism for the 
pulsations of HD~134214 is not clear, but further investigation of these modes may be
a probe of the atmospheric structure in this magnetic chemically peculiar star.
\end{abstract}


\section{Introduction}

HD~134214 (HI Lib) is a cool Ap (CP2) star with a spectral type of F2 Sr Eu Cr 
\citep{ren09}. \citet{kreidl85} and \citet{kk86} discovered light variations in this 
star with a period of 5.65\,min, confirming HD~134214 as a member of class of rapidly 
oscillating Ap (roAp) stars. The roAp class, established by \citet{ku82}, now consists of 
about 40 known members \citep[see for a list of members][]{ku06}. The 5.65-min period of 
HD~134214 is the shortest known among the roAp stars.  Only one frequency was detected in 
the first observations, in later groundbased photometry \citep{kr94} and in 10 hours of 
photometric monitoring by the MOST satellite in 2006 \citep{ca06}. 

The monoperiodicity of HD~134214 was rejected after 2 hours of time-resolved spectroscopy
by \citet*{ku06b} who detected 5 additional frequencies in the radial velocities of some 
lines of rare earth elements.  Soon after, 8.8\,h of spectroscopy by \citet{ku07} led 
to the identification of 2 more frequencies. Furthermore, new photometry they obtained 
revealed a second significant frequency in light. \citet{ku07} found that the photometric 
amplitude ratio of the secondary to the principal frequencies ($\approx 0.17$) is similar 
to the ratio in radial velocity amplitudes.  Recently, \citet{gru11} detected 10 
independent oscillation frequencies in HD~134214 from a 26-d (2008 April - May) 
photometric time series obtained by MOST, confirming all 8 previously 
reported frequencies. 
HD~134214 is now one of the richest multi-periodic roAp stars known, 
comparable to HD~24712 
(HR 1217) \citep{ku05}, HD~101065 (Przybylski's star) \citep{mkr08}, and
HD~60435 \citep*{mat87}.

The pulsations observed in roAp stars are p-modes of low degree and high radial order,
affected by globally coherent magnetic fields of strengths $1 \leq B \leq 25$\,kG. The 
periods range from 5.6 to 21\,min.  The amplitudes of many roAp pulsation modes are 
modulated with the star's rotation period.  This has been explained by the Oblique 
Pulsator Model proposed by \citet{ku82}, in which the pulsation is axisymmetric with 
respect to the dipole magnetic axis, which is itself obliquely inclined to the stellar 
rotation axis by an angle $\beta$. However, no amplitude modulation has been observed in 
HD~134214 in any of the photometric or spectroscopic data obtained to date. This may 
indicate that the magnetic obliquity is small ($\beta \approx 0^{\circ}$), or that the 
rotation axis is nearly coincident with the line of sight ($i \approx 0^{\circ}$) 
\citep{kk86}.

How does the strong magnetic field of an roAp star affect its pulsations? Lorentz forces 
on the moving plasma in the stellar atmosphere mean that the angular dependence of the 
eigenfunction of a nonradial mode cannot be described by a single spherical harmonic.
In addition, p-mode (acoustic) oscillations couple with magnetic oscillations in the 
outer layers where the Alfv\'en velocity $(v_{\rm A})$ is comparable to the sound speed  
$(c_{\rm s})$. Due to this coupling, part of the acoustic oscillation energy is converted 
to magnetic oscillations which propagate inward as slow waves.  These slow waves are 
expected to be dissipated deep in the interior of the star because of their very short
spatial wavelengths \citep{rs83}; i.e., magnetic coupling damps the p-mode oscillations. 
Including these effects, the properties of adiabatic p-mode oscillations in the presence 
of dipole magnetic fields were first investigated by \citet{dg96}, \citet{big00}, 
\citet{cg00}, and \citet{sg04}.  The results from different methods qualitatively agree 
with each other \citep{sa08}.  The oscillation frequency of a p mode generally increases 
as the strength of the magnetic field increases, because the phase velocity of 
magneto-acoustic waves $\sqrt{c_{\rm s}^2+v_{\rm  A}^2}$ increases. The gradual frequency 
increase is interrupted occasionally by an abrupt fall of a few tens of $\mu$Hz, when the 
damping due to slow-wave dissipation reaches a peak \citep{cg00,sg04}. 

\citet{cu06} modified the method of \citet{cg00} to make it possible to 
treat quadrupole magnetic fields, while \citet{sa05} extended the method 
of \citet{sg04} to nonadiabatic oscillations.  Furthermore, \citet{bd02} 
and \citet{bk11} found that pulsation axis does not necessary align with 
the magnetic axis if rotational effects are taken into account. 
Theoretical pulsation analyses including magnetic effects have been applied 
to several roAp stars to fit observed frequencies; for example, 
$\gamma$ Equ \citep{gru08}, 10~Aql \citep{hub08}, 
HD~101065 \citep{mkr08},  HD~24712 \citep*[=HR~1217][]{sa10}, and three roAp stars 
recently found by the Kepler satellite \citep{bal11a,bal11b,ku11}.

In this paper, we compare oscillation frequencies of HD~134214 obtained 
by \citet{gru11} from 2008 MOST photometry to theoretical frequencies of 
axisymmetric low-degree modes calculated by the method of \citet{sa05}, 
including the effect of dipole magnetic fields.

\section{Fundamental parameters of HD~134214}\label{sec:param}

In the literature, we note two recent sets of estimates of luminosity and effective 
temperature: $\log L/{\rm L}_\odot = 0.882 \pm 0.083$ and $\log T_{\rm eff} = 3.849 \pm 0.018$ 
by \citet*{hubrig00}, and $\log L/{\rm L}_\odot=0.85\pm0.07$ and $\log T_{\rm eff}=3.858\pm0.012$
by \citet{koc06}.  In both cases, the Hipparcos parallax was used to derive luminosity,
and  photometric indices of the Str\"omgren and the Geneva system to obtain $T_{\rm eff}$. 
Also, \citet*{rya08} obtained 
$T_{\rm eff} = 7315$ ($\log T_{\rm eff}=3.864$) and $\log g = 4.45$ from Str\"omgren 
photometry. 

\begin{figure}
\epsfig{file=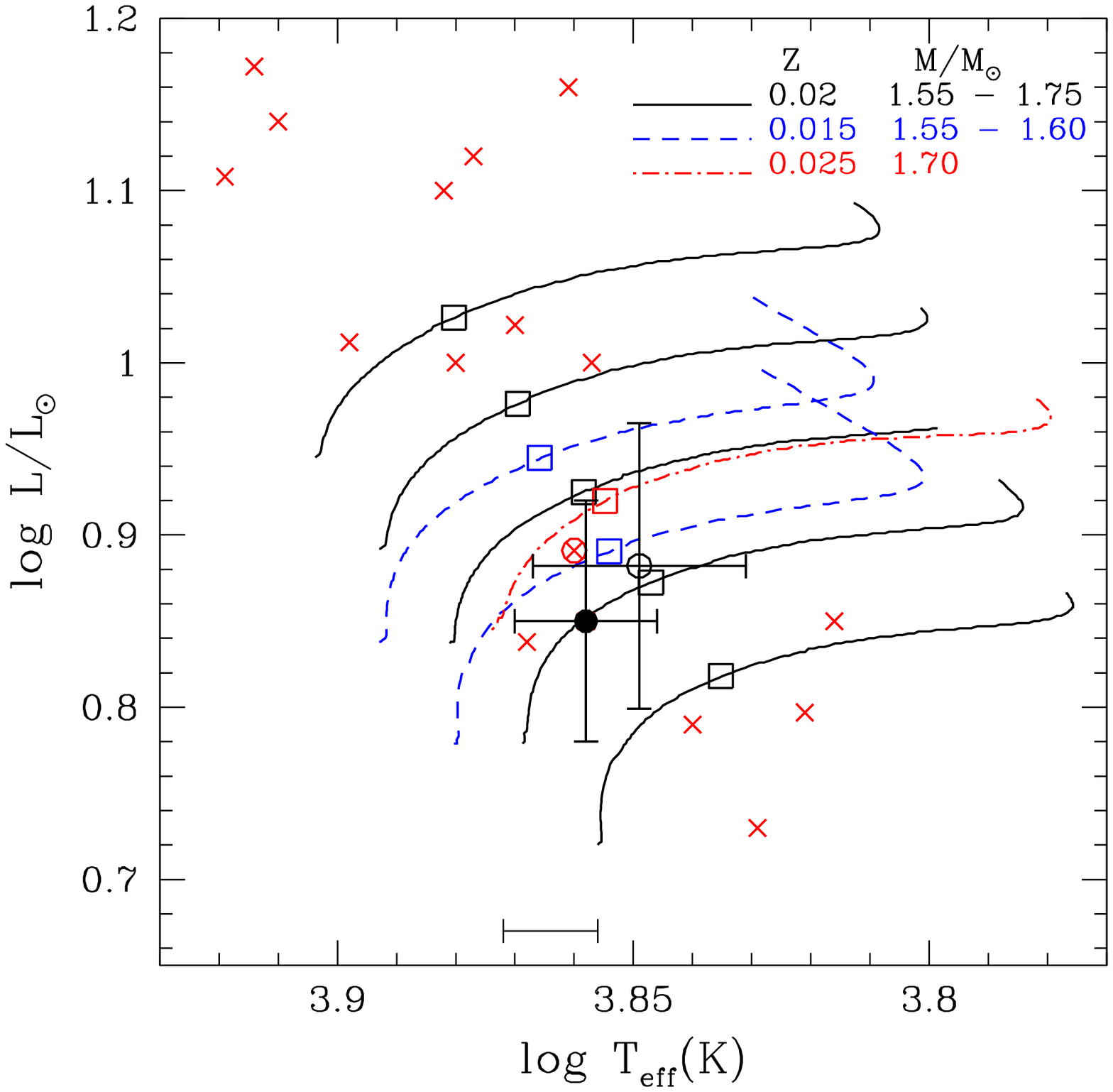,width=\columnwidth}
\caption{HR Diagram showing evolutionary tracks and the positions of HD~134214 
estimated by \citet{koc06} and \citet{hubrig00}; 
filled and open circles (with error bars), respectively. 
The range of $T_{\rm eff}$ near the bottom of the diagram is from \citet{rya08}.  Crosses 
show positions of other roAp stars; the circled cross represents HD~24712 (HR~1217).  The 
open square on each evolutionary track indicates the model whose oscillation frequencies 
best fit the observed frequencies of HD~134214. 
}
\label{fig:hrd}
\end{figure}

The possible locations of HD~134214 in the HR diagram (HRD) set 
by these estimates are shown in Fig.~\ref{fig:hrd}; 
filled and open circles with error bars are for the \citet{koc06} 
and \citet{hubrig00} estimates, respectively, and the range of $T_{\rm eff}$ obtained by 
\citet{rya08} is shown near the bottom of the diagram. Also shown in this figure are the 
positions of other roAp stars (crosses) adopted mainly from \citet{bal01}
 and \citet{koc06}.
This figure indicates that HD~134214 is a relatively cool roAp star and its location in
the HRD is similar to that of the prototypical roAp star HD 24712 (= HR 1217). 
Not only their positions in the HRD are similar, but also their pulsational 
properties \citep{gru11}. 
We will return to this similarity later in the paper.

\citet{mat97} obtained a mean surface magnetic field $\langle B_{\rm s}\rangle$ 
of 3.1\,kG from the Zeeman splitting pattern, but found no measurable longitudinal 
fields in HD~134214.
They also found a slight modulation of the mean field strength with a period of 
$4.15$ days, although \citet{ad00} found no long-term variation in the mean light. 
If the periodicity indicates the rotation period of HD~134214, it corresponds 
to an equatorial velocity of $\sim20$~km~s$^{-1}$.  
Combined with the projected rotation velocity $v_{\rm e}\sin i = 2$\,km\,s$^{-1}$ 
obtained by \citet{rya08}, we find the inclination angle to be as small as 
$\sim6^{\circ}$.  Such a nearly pole-on condition in HD 134214 was also inferred 
from the absence of pulsation amplitude modulations (and no rotational frequency 
splittings) by \citet{kr94}.

\section{Stellar and pulsation models}

We have generated unperturbed stellar models for our pulsation analysis from evolution 
models calculated by a Henyey-type code for a mass range 
of $1.55 \le M/{\rm M}_\odot \le 1.75$, 
using OPAL opacity tables \citep{opal95} supplemented with low-temperature tables by 
\citet{alex94}. Some of the calculated evolutionary tracks are shown 
in Fig.~\ref{fig:hrd}.

For the outermost atmospheric layers we have employed the standard $T-\tau$ 
relation given 
in \citet{ss85}. In some models, helium is assumed to be depleted above the first He 
ionization zone (appropriate for the stratified atmosphere of an Ap star) in the same way 
as described by \citet{sa10}.  We have suppressed envelope convection in majority of the 
models, assuming that a strong magnetic field should inhibit convective 
instability in the envelope.  
For those models where we did not suppress envelope convection, we used the 
mixing-length theory formulated by \citet*{heny65} with a mixing length of 1.5 pressure 
scale heights. Core convection is always included, but with no overshooting.  
Table~\ref{tab:models} lists the assumed parameters adopted for each
series of evolutionary models, in which `Y' and `N' mean `included' and `not included', 
respectively.  

\begin{table}
\caption{Unperturbed model parameters}
\label{tab:models}
\begin{tabular}{@{}lccccc}
\hline
Model & $M/{\rm M}_\odot$ & $X$ & $Z$  &  helium   &  envelope  \\
name  &             &     &      & depletion & convection \\
\hline
D155    & 1.55 & 0.700 & 0.020  & Y & N  \\
D160    & 1.60 & 0.700 & 0.020  & Y & N  \\
D165    & 1.65 & 0.700 & 0.020  & Y & N  \\
D170    & 1.70 & 0.700 & 0.020  & Y & N  \\ 
D175    & 1.75 & 0.700 & 0.020  & Y & N  \\ 
H170    & 1.70 & 0.700 & 0.020  & N & N  \\
H160C   & 1.60 & 0.700 & 0.020  & N & Y  \\
H165C   & 1.65 & 0.700 & 0.020  & N & Y  \\
H170C   & 1.70 & 0.700 & 0.020  & N & Y  \\
D155Z15 & 1.55 & 0.705 & 0.015  & Y & N  \\
D160Z15 & 1.60 & 0.705 & 0.015  & Y & N  \\
D170Z25 & 1.70 & 0.695 & 0.025  & Y & N  \\
\hline
\end{tabular}
\end{table}

We have calculated nonadiabatic frequencies for axisymmetric p-modes in the presence 
of dipole magnetic fields (but without rotation) using the method described by 
\citet{sa05} and \citet{sg04}, where perturbations of magnetic fields are included 
by adopting the ideal MHD approximation.  The pulsation axis is assumed to be aligned 
with the magnetic axis. Frequencies are obtained for field strengths at intervals of 
0.1~kG in the range $2\,{\rm kG} \le B_{\rm p} \le 7\,{\rm kG}$ (where $B_{\rm p}$ = 
polar field strength). Since the angular dependence of the eigenfunction of a 
pulsation mode in the presence of a magnetic field deviates from a single spherical 
harmonic, we express it by a sum of terms proportional to Legendre functions 
$P_\ell(\cos\theta)$ (or $Y_\ell^{m=0}$, where $\theta$ is co-latitude measured from the
 pulsation axis) with $\ell = 0,2,4, \ldots$ for an even mode, or $\ell = 1,3, 5, 
 \ldots$ for an odd mode; i.e., the kinetic energy of a pulsation mode is distributed 
among components with different degrees $\ell$.  To designate the angular dependence 
of a mode we use ${\ell}_{\rm m}$ which is equal to the degree $\ell$ of the 
component having the largest fraction of kinetic energy among the expanded terms of 
the mode.
In most cases, we have 
employed twelve $\ell$ values to express an eigenfunction. When the expansion did not 
converge sufficiently after 12 terms, up to 14 Legendre functions were used.

It is not easy to estimate a mean error for theoretical frequencies, because 
uncertainty depends on the frequency and $B_{\rm p}$.
When the convergence of expansion is reasonably good (i.e., the ratio of the
kinetic energy of the last component to that of the main component is 
less than 10\,percent), 
the differences between the frequencies obtained using 12 and 14 Legendre functions 
are less than $0.5\,\mu$Hz. 
But among the cases with a mediocre convergence with the above ratio 
being 10-30\,percent,
the differences in some cases are $\sim\,1\mu$Hz 
(in rare cases the difference  can be up to $1.5\,\mu$Hz).
From these properties, we consider the mean error of the theoretical frequencies 
due to the truncation of the expansion to be $\sim 0.5\,\mu$Hz. 
In addition, we expect errors from using spherically symmetric equilibrium models 
despite strong magnetic fields.
Although it is not clear how seriously the magnetic non-sphericity would affect 
the frequencies of high-order p-modes,  we assume that the uncertainty in 
theoretical frequencies would be $\sim 1\,\mu$Hz.

Since the observed pulsation frequencies of HD~134214 are all above the acoustic 
cut-off frequency (as in the case of HD~24712), we have applied the running wave 
outer boundary condition described in \citet{sa10}.  In the pulsation analysis for 
models that include envelope convection, we have used a frozen convection approximation 
for the divergence of convective energy flux. The perturbation of radiative flux is 
treated in the same way as done by \citet{sa10}, adopting the \citet{un66} theory for 
the time-dependent Eddington approximation.

\section{Frequency fittings} \label{sec:fit}

Table~\ref{freqid} lists frequencies and photometric amplitudes obtained by 
\citet{gru11} from MOST photometry, and radial velocity amplitudes of Nd III lines 
obtained by \citet{ku07}.  The first column contains the designations of frequencies 
by \citet{gru11}, and the second column lists the corresponding designations by 
\citet{ku07}. No frequencies corresponding to $\nu_9$ and $\nu_{10}$ were found by 
\citet{ku07}.  The ranking of the photometric amplitudes is very different than
that of the radial velocity amplitudes (except for $\nu_1$) which might be related 
to differences in mode properties and/or long-time variations in pulsation amplitudes 
\citep{ku07}. The last column of the table indicates ${\ell}_m$ for each frequency 
identified in our best models (see below).

\begin{table}
\caption{Observed frequencies and amplitudes of HD 134214}
\label{freqid}
\begin{tabular}{@{}lccccc}
\hline
ID1$^{\rm a}$ & ID2$^{\rm b}$ & freq$^{\rm a}$ & Amp$^{\rm a}$ & Amp(Nd III)$^{\rm b}$ 
& $\ell_{\rm m}$ \\
       &       & (mHz) & (mmag) & (m s$^{-1}$) & \\
\hline
$\nu_1$    & $\nu_{{\rm k}1}$ & 2.9495 & 1.820 &   374.3  &  1  \\
$\nu_2$    & $\nu_{{\rm k}6}$ & 2.9157 & 0.174 &    22.1  &  2  \\
$\nu_3$    & $\nu_{{\rm k}2}$ & 2.7795 & 0.157 &    60.6  &  2  \\
$\nu_4$    & $\nu_{{\rm k}7}$ & 2.9833 & 0.116 &    22.4  &  2  \\
$\nu_5$    & $\nu_{{\rm k}3}$ & 2.6470 & 0.063 &    36.4  &  2  \\
$\nu_6$    & $\nu_{{\rm k}5}$ & 2.7220 & 0.061 &    38.8  &  1  \\
$\nu_7$    & $\nu_{{\rm k}8}$ & 2.8053 & 0.054 &    19.4  &  0  \\
$\nu_8$    & $\nu_{{\rm k}4}$ & 2.8416 & 0.051 &    22.5  &  3  \\
$\nu_9$    & ----             & 2.6872 & 0.048 &    ----  &  1? \\
$\nu_{10}$ & ----             & 2.6998 & 0.044 &    ----  &  3? \\     
\hline
\end{tabular}\\
$^{\rm a}$Adopted from \citet{gru11}.\\
$^{\rm b}$Adopted from \citet{ku07}.
\end{table}

For the fittings to be discussed in this section, 
we adopt only the eight confirmed frequencies ($\nu_1, \nu_2, \ldots, \nu_8$)
common to both the photometric and spectroscopic analyses. 
We omit the two frequencies having lowest amplitudes ($\nu_9$ and $\nu_{10}$), 
allowing for the possibility that they could be modes of higher degree ($\ell_m\ge 4$). 
We will discuss later the result of fits including all 10 frequencies.

For each model sequence listed in Table~\ref{tab:models}, we have looked for 
a model whose frequencies of low-degree ($\ell_m \le 3$) axisymmetric modes best 
fit the eight frequencies ($\nu_1, \nu_2, \ldots, \nu_8$).
The ``goodness" of a model fit is measured by the mean deviation (MD) from the 
eight frequencies, where we have adopted the same weight for all eight, because 
the uncertainties in observed frequency values ($0.05\,\mu$Hz) are much 
smaller than theoretical ones ($\sim 1\,\mu$Hz).  The theoretical frequencies have 
been interpolated with respect to radius along each evolutionary path to find 
the best radius, but no interpolation is performed with respect to $B_{\rm p}$ 
because frequency may change discontinuously as a function of $B_{\rm p}$. 
(Recall that model frequencies are calculated at intervals of 0.1\,kG.)

\begin{table}
\caption{Best fitting models}
\label{tab:bestmodels}
\begin{tabular}{@{}lccccc}
\hline
Model    & $\log R$ & $\log T_{\rm eff}$ & $\log L$ & $B_p$(kG) & MD($\mu$Hz)
 \\
\hline
D155     &  0.2623 & 3.8352 & 0.8180 & 4.8 &  1.31 \\
D160     &  0.2659 & 3.8470 & 0.8723 & 4.4 &  1.29 \\
D165     &  0.2694 & 3.8584 & 0.9250 & 4.1 &  1.23 \\
D170     &  0.2728 & 3.8695 & 0.9761 & 3.8 &  1.33 \\
D175     &  0.2761 & 3.8803 & 1.0262 & 3.9 &  1.29 \\
H170     &  0.2740 & 3.8691 & 0.9769 & 4.3 &  1.32 \\
H160C    &  0.2707 & 3.8453 & 0.8753 & 7.0 &  1.22 \\
H165C    &  0.2729 & 3.8572 & 0.9271 & 5.9 &  1.20 \\
H170C    &  0.2747 & 3.8689 & 0.9773 & 4.6 &  1.28 \\
D160Z15  &  0.2645 & 3.8658 & 0.9448 & 4.3 &  1.32 \\
D170Z25  &  0.2740 & 3.8548 & 0.9197 & 4.3 &  1.26 \\
\hline
\end{tabular}
\end{table}

\begin{figure}
\epsfig{file=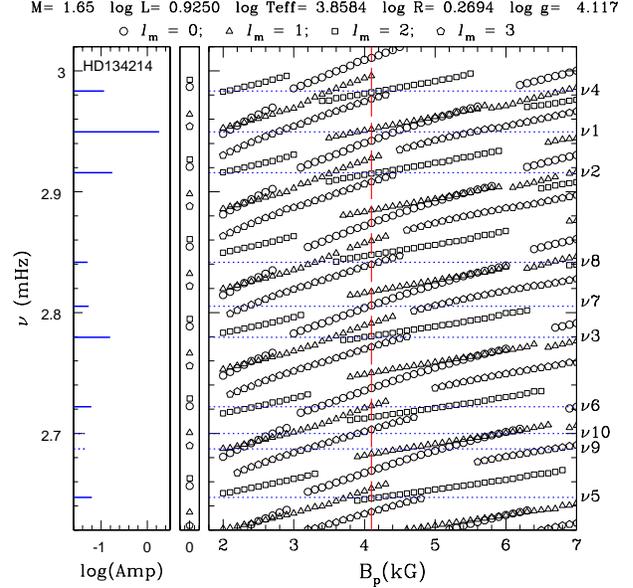,width=\columnwidth}
\caption{
Left panel: Observed pulsation frequencies (mHz) and amplitudes (mmag) of HD 134214
obtained by \citet{gru11} from the 2008 MOST photometry. Note that the amplitudes 
are plotted as log values to accommodate the amplitude of the principal frequency 
(2.950 mHz), which is an order of magnitude larger than any other.  Right panel: 
Pulsation frequencies of low-degree modes ($\ell_{\rm m} \le 3$) as a function
of polar magnetic strength $B_{\rm p}$ for the best fitting model of mass $1.65\,{\rm M}_\odot$, 
whose parameters are shown at the top of the diagram.  Horizontal dotted 
lines indicate observed frequencies, whose designations are given along the right 
vertical axis. The model well reproduces the observed frequencies $\nu_1, \nu_2, 
\ldots, \nu_8$ at $B_{\rm p}=4.1$\,kG (vertical dashed line). Middle (narrow) panel:
Model frequencies with no magnetic field ($B_{\rm p} = 0$).
}
\label{fig:nubp}
\end{figure}

Table~\ref{tab:bestmodels} lists parameters of the best fitting models we have
found for various model sequences; the mean deviations (MD) are given in the 
last column. The loci of these best fitting models in the HRD are shown by open 
squares in Fig.~\ref{fig:hrd}. The mean deviations of the best models for
different masses vary only slightly.  A model with $1.65\,{\rm M}_\odot$ best reproduces 
the eight observed frequencies at polar magnetic field strength $B_{\rm p} = 4.1$\,kG 
among the D sequences, in which a standard composition $(X,Z) = (0.70,0.02)$ 
is adopted, envelope convection is suppressed, and helium above the first He 
ionisation zone is depleted. The required magnetic field strength is plausible, 
given the mean modulus (3.1\,kG) for this star obtained spectroscopically 
\citep{mat97}. Fig.~\ref{fig:nubp} compares the theoretical frequencies of the 
best model as a function of $B_{\rm p}$ with the observed frequencies; the closest 
agreement occurs for $B_{\rm p} = 4.1$\,kG (vertical dashed line). Generally, pulsation 
frequencies increase as the magnetic field strength increases, roughly preserving 
the large separation.  However, since the dependence of frequencies on $B_{\rm p}$ 
differs slightly for different $\ell_{\rm m}$, the relative loci of sequences for 
different degrees $\ell_{\rm m}$ differ considerably from the case with no magnetic field.

\begin{figure}
\epsfig{file=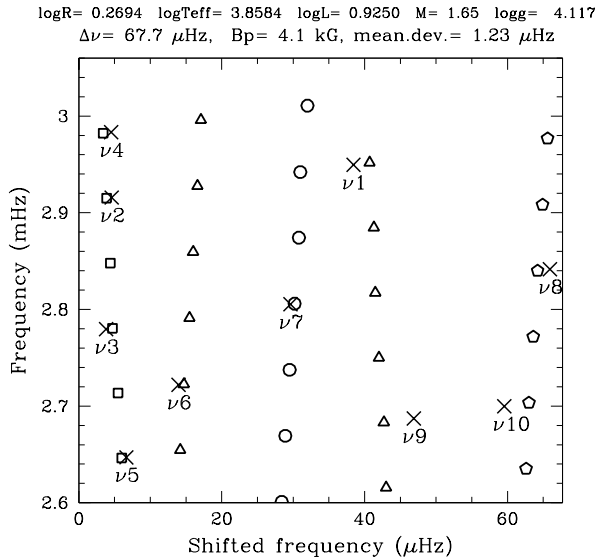,width=\columnwidth}
\caption{Echelle diagram (folded modulo $67.7\,\mu$Hz) for the best fitting 
model with polar magnetic field strength $B_{\rm p} = 4.1$\,kG of the 
D165 ($1.65\,{\rm M}_\odot$) model sequence. 
Crosses are observed frequencies whose designations are taken from \citet{gru11}. 
Open circles, triangles, squares, and pentagons represent theoretical 
frequencies for $\ell_{\rm m} = 0, 1, 2,$ and $3$, respectively. 
Mean deviation MD is shown for $\nu_1, \nu_2, \ldots, \nu_8$.
}
\label{fig:echelle}
\end{figure}

The slow increase in frequency with $B_{\rm p}$ is interrupted occasionally by a 
sudden decrease of a few tens of $\mu$Hz; this behaviour was first recognised by 
\citet{cg00}. Fig.~\ref{fig:nubp} shows that, in such a transition range, two 
distorted dipole modes appear.  This is why two dipole modes appear within a large 
frequency separation in the echelle diagram shown in Fig.~\ref{fig:echelle}. These 
two deformed dipole modes have $\ell_{\rm m} = 1$ by definition, but have different 
distributions of kinetic energy among components of $\ell = 3, 5. \ldots$, and 
hence have different angular dependences. Such additional modes are needed to 
explain the frequency spectrum of the roAp star HD 101065 \citep{mkr08}.

The theoretical frequencies of the model with $B_{\rm p} = 4.1$\,kG are compared in 
detail as an echelle diagram in Fig.~\ref{fig:echelle}. In this diagram, frequencies 
are folded modulo $67.7\,\mu$Hz according to the large frequency separation obtained 
by \citet{gru11} from their observed frequencies of HD~134214. The large separation 
of the model is consistent with the observed one, although the former slightly 
depends on degree $\ell_{\rm m}$ due to the fact that the magnetic effects also 
depend on $\ell_{\rm m}$.  We find reasonably good agreement between theoretical and 
observed frequencies.  The mean deviation MD of this model ($1.23\,\mu$Hz) seems 
to be slightly larger than the uncertainty expected for theoretical frequencies.  
A non-dipole component of the magnetic field could account for the discrepancy.

We identify the principal frequency ($\nu_1=2.950$\,mHz) as a distorted dipole mode, 
while frequencies $\nu_2, \nu_3, \nu_4,$ and $\nu_5$ are identified as 
$\ell_{\rm m}=2$ modes.  
We note that the frequencies $\nu_2, \nu_3,$ and $\nu_4$, have
similar amplitudes (next largest after the principal frequency) and, hence, appear 
to share the same degree. Mode identifications in the other best fitting models 
listed in Table~\ref{tab:bestmodels}  are similar to those by D165. They are listed 
in the last column of Table~\ref{freqid}.

Comparing the best models of the  D170 and H170 sequences, we see that helium 
depletion in the upper atmosphere does not affect the ``goodness" 
of the best model. 
The magnetic field strength of the best fit from the helium-depleted model 
sequence (D170) is smaller than the model without helium depletion (H170). 
A stronger magnetic field is needed for the latter model to have the same 
magnetic effect, because the mean density of the model without helium 
depletion is larger than the helium-depleted model at the same $L$ and 
$T_{\rm eff}$, due to the lower opacity and higher mean-molecular weight. 

It can be seen in Table~\ref{tab:bestmodels} that including convective energy 
transport in the envelope convection zone (H160C, H165C, and H170C) slightly 
improves the quality of the fits but only by adopting stronger magnetic fields.
The improvement, however, seems too weak to claim the presence of efficient 
convection in the polar regions of HD~134214.  
The necessity of the stronger fields is due to higher mean density in the 
outer envelope as discussed above.
In this case, the higher density is caused by, 
in addition to the homogeneous helium abundance in the envelope, 
a decrease in the temperature gradient brought by convective energy transport. 
Changing metallicity does not significantly change the quality of fits.

\section{Mode properties}

\begin{figure}
\epsfig{file=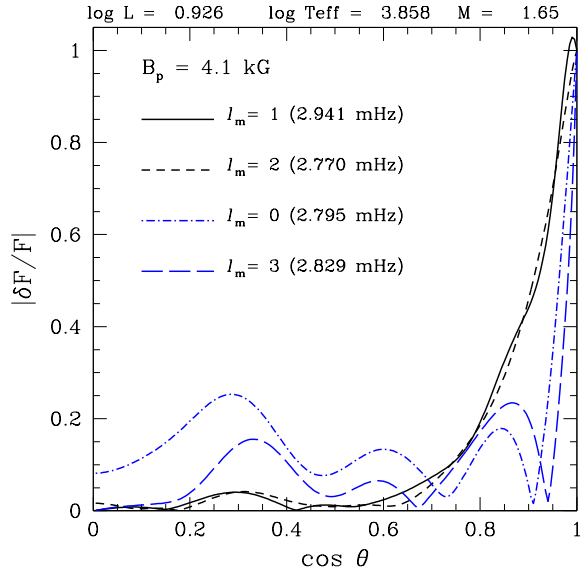,width=\columnwidth}
\caption{The latitudinal amplitude distribution of flux perturbation as a 
function of co-latitude $\theta$ on the stellar surface of a model with 
M = $1.65\,{\rm M}_\odot$ and $B_{\rm p} = 4.1\,$kG. 
The solid line is for a deformed dipole mode which fits the 
principal frequency of HD~134214, the short dashed line for a $\ell_m=2$  
mode fitted to $\nu_3 (=\nu_{{\rm k}2})$,
the dash-dotted line for a
$\ell_{\rm m}=0$ mode fitted to $\nu_7$, and the long dashed line
for a $\ell_m=3$ mode fitted to $\nu_8$.
}
\label{fig:lat_amp}
\end{figure}

In the best fitting models discussed above, the principal frequency corresponds to 
a deformed dipole mode, while $\nu_2 \ldots \nu_5$ match $\ell_m=2$ modes 
whose amplitudes are very much smaller than that of the principal one (see 
Table~\ref{freqid}). 
One could speculate that the large amplitude difference might be a 
difference in visibility due to different angular dependences of the amplitudes 
on the stellar surface between modes of $\ell_{\rm m} = 1$ and $\ell_{\rm m} = 2$. 
However, a glance at Fig.~\ref{fig:lat_amp} demonstrates that this is not the case.
This figure shows the latitudinal dependences of flux perturbations of the dipole 
mode (solid line) and one of the $\ell_{\rm m}=2$ modes, $\nu_3$ 
(short dashed line) for $B_{\rm p}=4.1$\,kG. 
The latitudinal dependences are similar to each other, 
although without a magnetic field, a dipole mode should vary as $\cos\theta$, 
quite different from the expected $\ell=2$ 
dependence (${1\over2}(3\cos^2\theta-1)$).
The strong magnetic field suppresses pulsation 
amplitudes at low latitudes because gas cannot cross field lines, so that large 
amplitudes are strongly confined to the polar regions. 

The only appreciable difference between the two modes is the parity 
with respect to the equatorial plane; 
$\ell_{\rm m}=1$ modes are odd and $\ell_{\rm m}=2$ modes are even. 
For the principal frequency to be identified as a deformed dipole mode, 
the angle between the magnetic axis and line-of-sight should be small 
so that the contribution from the opposite hemisphere is small.  
Combining this fact with the requirement of a small angle between 
the rotation axis and the line of sight (Sec.\ref{sec:param}) indicates 
the obliquity angle $\beta$ of HD~134214 must be small too.  
With a small angle between the line of sight and the pulsation axis, 
there is little difference in the visibility between $\ell_m = 1$ 
and $\ell_m = 2$ modes. This is contrary to the conjecture of \citet{gru11} 
that the large amplitude difference between $\nu_1$ and other frequencies 
is due to a large difference in their visibilities.

Fig.~\ref{fig:lat_amp} also shows the latitudinal distributions of 
flux perturbation for $\ell_{\rm m}=0$ and 3 modes which are fitted 
to $\nu_7$ and $\nu_8$, respectively.
In contrast to the cases of $\ell_{\rm m}=1$ and 2, the flux amplitudes of 
the $\ell_{\rm m}=0$ and 3 modes show wavy variations as a function of latitude
with quasi-nodal lines which would reduce the visibility of the modes.
(Note that the $\ell_m=0$ mode is significantly deformed from spherical
symmetry due to the contributions from high $\ell$ components.)
This is consistent with the fact that $\nu_7$ and $\nu_8$ have smallest 
amplitudes among the eight frequencies adopted here.

\begin{figure}
\epsfig{file=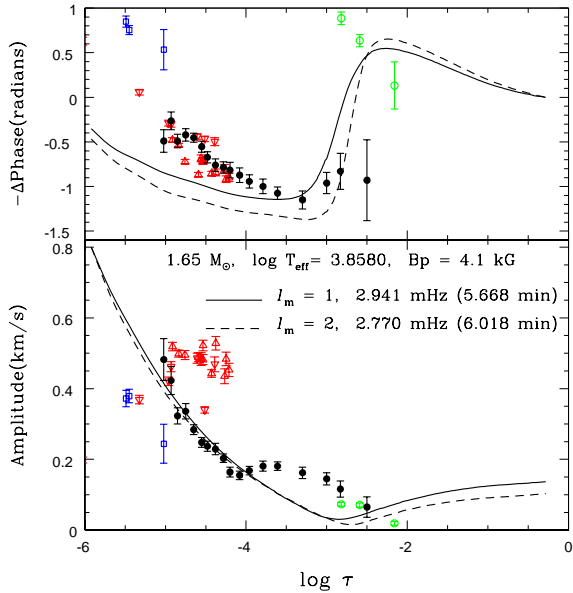,width=\columnwidth}
\caption{Pulsation phase (top panel) and radial velocity amplitude 
(bottom panel) variations in the atmosphere of a $1.65\,{\rm M}_\odot$ 
model close to the best 
fitting model with $B_{\rm p} = 4.1$\,kG, for $\nu_1$ (solid lines) and 
$\nu_3 ( = \nu_{{\rm k}2})$ (dashed lines).
The radial velocity amplitude is set to be 0.8\,km s$^{-1}$ at $\log \tau = -6$,
while the phase is set to be zero in the photosphere, where $\tau$ is 
the optical depth calculated with the Rosseland mean opacity. 
In this diagram, the pulsation (and hence 
magnetic) axis is assumed to be along the line of sight.
Also shown are radial velocity data of HD 134214  taken from \citet{rya07}; 
{\Large\textbullet}=bisector of H$\alpha$ core,  ${\bigcirc}$=Y II, 
$\bigtriangleup$=Nd II, $\bigtriangledown$=Nd III, and $\sq$=Pr III, 
where the optical depth at each H$\alpha$ line depth and the formation 
depths of the other lines are assumed to be the same as in the case of HD 24712.
}
\label{fig:tau_amp}
\end{figure}

Fig.~\ref{fig:tau_amp} shows runs of phase delay and amplitude of radial 
velocity variations in the atmosphere of a model which has parameters 
close to those of the best (interpolated) model.  
In this figure, the pulsation axis (and hence the magnetic axis) is 
assumed to be aligned with the line of sight. 
The properties are not very sensitive to the 
obliquity of pulsation and magnetic axes; we see very similar model behaviours 
if we adopt, for example, an obliquity of $30^\circ$ instead of $0^\circ$.  
Fig.~\ref{fig:tau_amp} reveals a quasi-node around $\log\tau \approx -3$ 
for both modes, where $\tau$ is again the optical depth calculated with the 
Rosseland mean opacity.  
Above the quasi-node,  the amplitude and phase delay increase outward, 
indicating the presence of outwardly running wave components.  
This is consistent with the fact that both frequencies are above the critical 
acoustic cut-off.  
We note that the quasi-node is not the ``false node'' discussed by 
\citet{sc11}, because it appears in the eigenfunction.

Also plotted in Fig.~\ref{fig:tau_amp} are data of radial velocity/phase 
measurements from H$\alpha$ bisector and lines of some rare earth elements 
of HD 134214 adopted from \citet{rya07}, where the optical depth 
at each H$\alpha$ line depth and the formation depths of 
the other lines are assumed to be the same as in the case of HD 24712.
This figure shows that the observational data qualitatively agree with 
the model; amplitude and phase-delay increases outward in the outermost layers, 
which also agree with the spectroscopic analysis by \citet{ku07}.
However, the observed variation of phase delay in the outermost layers is 
considerably steeper than the model prediction.  
The discrepancy probably indicates that the sound speed (or temperature) in the 
upper atmosphere of HD~134214 is lower than in the model. 
This is consistent with the actual temperature gradient being steeper
%This might indicate in turn the actual temperature gradient to be steeper 
due to a strong blanketing effect which is not included in our models.

The radial velocity amplitudes measured by Nd II/III tend to be larger 
than those of H$\alpha$, while the amplitudes of Pr III lines tend to be 
smaller than those of H$\alpha$. 
These differences can be attributed to the difference in the distributions of 
the elements Nd and Pr on the stellar surface. 
If the depths of line formation are similar to HD~24712, as we have surmised,
then Figs.~\ref{fig:lat_amp}~and~\ref{fig:tau_amp} suggest that Nd is 
distributed at a higher pulsational latitude than Pr.

At the quasi-node, the pulsation phase of the principal frequency 
$\nu_1$ changes dramatically by about 1.8 radians. 
The presence of the phase shift at $\tau\approx -3$ is supported 
observationally by the phase shift from the deep part of H$\alpha$ 
phase to the phases of Y II (Fig.~\ref{fig:tau_amp}).

The runs of phase and amplitude in the atmosphere for 
$\nu_3 (=\nu_{{\rm k}2}; \ell_{\rm m} = 2)$ shown in dashed lines in 
Fig.~\ref{fig:tau_amp} are very similar to those for the principal frequency 
($\ell_{\rm m} = 1$; solid lines). This agrees with the finding by 
\citet{ku07} that the amplitude ratio of the two frequencies 
($\nu_1$ and $\nu_3$) from the radial velocity variations 
(weighted toward the outermost layers) is similar to that 
measured for the light variations (which occur in the photosphere).

The similarity in the latitudinal and depth dependences between the 
principal mode and the secondary mode suggests that the huge amplitude 
difference is caused by a 
large difference in the excitation strengths of yet  unknown mechanism.

\section{Fitting all ten frequencies}

\begin{figure}
\epsfig{file=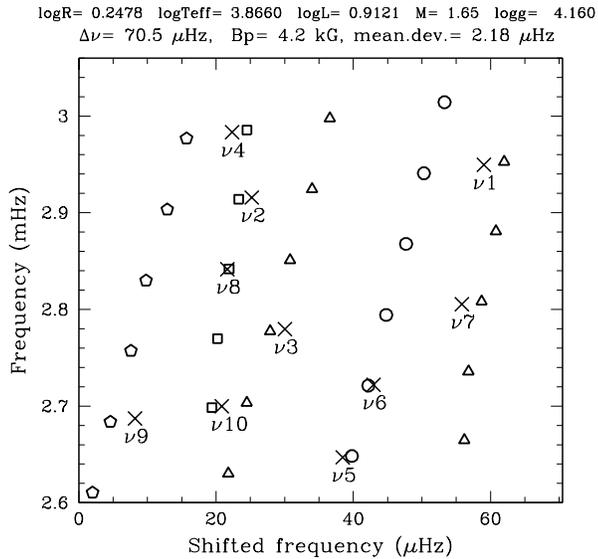,width=\columnwidth}
\caption{An echelle diagram (folded modulo $70.5\,\mu$Hz) for the model 
(among models of D165) which best fits (with a mean deviation of $2.18\,\mu$Hz) 
the ten observed pulsation frequencies of HD~134214. 
The model has a polar field strength of $B_{\rm p}= 4.2$\,kG. Symbols are the 
same as in Fig.~\ref{fig:echelle}.
}
\label{fig:fit10f}
\end{figure}

In the model fits described above, we omitted the two frequencies 
($\nu9$ and $\nu10$) with the lowest amplitudes because they might be due to 
high-degree ($\ell_m \ge 4$) modes for which no theoretical frequencies are available.  
Here we discuss fits performed for models in the 
D165 sequence assuming all ten observed frequencies are low-degree ($\ell_m \le 3$) modes.
Fig.~\ref{fig:fit10f} shows the echelle diagram for the best fitting model from the D165 
series.  The observed frequencies are best fitted with $B_{\rm p}=4.2$\,kG, resulting in a 
mean deviation MD = $2.18\,\mu$Hz. Compared to the best fitting D165 model for only eight 
frequencies (Table~\ref{tab:bestmodels}), this model is slightly hotter, but still within 
the uncertainty of the photometric effective temperature (Fig.~\ref{fig:hrd}).  
Note that in Fig.~\ref{fig:fit10f}, frequencies are folded with a separation of 
$70.6\,\mu$Hz, which is larger than what is used in Fig.~\ref{fig:echelle} 
($67.7\,\mu$Hz) because of the higher $T_{\rm eff}$.
In this model, $(\nu_4,\nu_2,\nu_8,\nu_{10})$ are identified as $\ell_m = 2$ modes rather 
than $(\nu_4,\nu_2,\nu_3,\nu_5)$. Comparing Fig.~\ref{fig:fit10f} (10-frequency fit) with
Fig.~\ref{fig:echelle} (8-frequency fit), the quality of fit is better in the latter case.
This is circumstantial evidence supporting the argument that $\nu_9$ and $\nu_{10}$ are 
modes of degree equal to or higher than 4.

\section{Discussion}

\subsection{Excitation mechanism}

The predicted frequencies and large separations of the best fitting model of each series 
are roughly consistent with the observed values for HD~134214.  However, there is one 
serious problem: all observed frequencies are above the acoustic cut-off frequencies of 
our models and all modes with appropriate frequencies are damped modes.

Pulsations of roAp stars are now generally considered to be excited by the
kappa-mechanism operating in the hydrogen ionization zone \citep{bal01,cu02}.  
Our models are located in the instability region of the HR Diagram obtained by 
\citet{cu02}.  
It turns out that p-mode pulsations of order up to $n\sim 20$ ($\nu \la 1.4$\,mHz) 
are actually excited by the kappa-mechanism in the best fitting D165 model 
($B_{\rm p} \sim 4-5$\,kG).  
But none of these predicted pulsations have been detected.

\citet{aud98} found that the cut-off frequency increases if the $T-\tau$ relations 
based on more physically accurate atmospheric models are used. 
For the $T_{\rm eff}$ range appropriate for HD~134214, however, even their models 
predict cut-off frequencies below about 2.7\,mHz, smaller than the principal 
frequency of 2.95\,mHz in HD~134214.  
Another way to increase the cut-off frequency is to introduce a temperature 
inversion in the atmosphere, as proposed by \citet{gsh98}. 
Unfortunately, the inversion necessary in this case must be
a few thousand degrees.  Even if we assume a reflective boundary condition, 
no modes having frequencies comparable to those of HD~134214 are excited.   
In addition, observations indicate the presence of outwardly running waves 
in the outermost layers (Fig.~\ref{fig:tau_amp}) 
do not support a reflective boundary.
Therefore, the outer boundary is probably not the cause of the 
excitation discrepancy.  

The same problem exists for another multi-periodic roAp star, HD~24712 (HR~1217), 
as discussed by \citet{sa10}. 
These two examples clearly indicate that an instability mechanism other than 
the kappa-mechanism must be operating, at least for the super-critical frequency 
pulsations.  
The mechanism must be intimately connected with the interaction with magnetic 
fields because no such high-frequency oscillations occur in A-F stars 
other than the roAp stars. 
The atmospheric layers with a super-adiabatic temperature gradient and a magnetic 
field are overstable in linear analyses with the Boussinesque approximation, 
where density perturbations are neglected except for buoyancy forces 
\citep[e.g.,][]{ch61}.  
Based on the fact that the periods of overstability for Ap stars with reasonable 
magnetic field strengths are comparable to the period range seen in roAp stars, 
\citet{shb83} predicted overstability to be the excitation 
mechanism for roAp stars. 
\citet{cox84} discussed some physical implications of overstability.  
Weaknesses of this concept were  identified by \citet{bal01}: 
as the analyses are local, it is not certain whether a global mode is excited, 
and even if global modes are excited it is not certain whether such a mode has any 
significant amplitude above the thin convective layer.  Further investigations 
of the roAp excitation mechanism are definitely needed.

\subsection{Similarities and dissimilarities to HD~24712}

\begin{table}
\caption{Comparing HD~134214 and HD~24712}
\label{tab:comparison}
\begin{tabular}{@{}l|c|cc}
\hline
 &  HD~134214 &  HD~24712 & refs. \\ 
\hline
Frequency range (mHz)                       & $2.65 - 2.98$  & $2.55 - 2.81$ & 1, 2, 3 \\ 
Large separation ($\mu$Hz)                  & 67.7           & 68            & 1, 2    \\
Small separation $\delta^{\rm a}$ ($\mu$Hz) & 0.02           & 0.5           & 1, 2    \\
Rotational splitting ($\mu$Hz)                  &  $-$           & 0.929         & 2       \\
\multicolumn{3}{@{}l}{Parameters of best-fit models}                         & 4, 5    \\
\hspace{0.5cm}$\log T_{\rm eff}$            & 3.8584         & 3.8585  \\
\hspace{0.5cm}$\log L/{\rm L}_\odot$              & 0.9250         & 0.9247  \\
\hspace{0.5cm} $M/{\rm M}_\odot$                  & 1.65           & 1.65    \\
\hspace{0.5cm} $B_{\rm p}$(kG)              & 4.1            & 4.9     \\
\hline
\end{tabular}
refs.:1=\citet{gru11}, 2=\citet{ku05}, 3=\citet{mkr05}, 4=this paper, 5=\citet{sa10}
$^{\rm a}\delta \equiv \nu(l_{\rm m}=1,n) - 0.5[\nu(l_{\rm m}=2,n-1)+ \nu(l_{\rm m}=2,n)]$
\end{table}

We have already mentioned several times in this paper the similarity of HD~134214 to
another roAp star HD~24712 (HR~1217), and Table~\ref{tab:comparison} compares a few 
properties of these two stars.  
The loci in the HR Diagram of the best fitting models for the two stars are 
nearly identical.  
The observed frequency ranges are above the acoustic cut-off frequencies of 
the models in both cases, and they are predicted to be damped. 
Both stars have similar large frequency separations and near zero small separations, 
$\delta$, which is difficult to reproduce with models. 
Our best model (shown in Fig.~\ref{fig:echelle})  gives $\delta\sim3\,\mu$Hz, 
which for HD~134214 is  much larger than the observed value of $0.02\,\mu$Hz.  
The same difficulty is reported for HD~24712 by \citet{sa10}. 
Our current investigation of HD~134214 makes it clear that the problems in 
modelling of HD~24712 are not unique to this star, and may be the same for other 
roAp stars with similar parameters.

For completeness, we note some dissimilarities between the two stars.  
HD~24712 exhibits clear rotational amplitude modulation and associated frequency 
splittings, while HD~134214 shows none.  
The amplitude of the principal frequency of HD~134214 (identified by us as 
an $\ell_{\rm m}=1$ mode) is very much larger than the other ones and has been nearly 
constant for a long time \citep{ku07,gru11}. On the other hand, HD~24712 has three main 
frequencies with comparable amplitudes; the relative amplitudes vary on a time scale of 
a few years \citep{ku05}.

\section{Conclusions}

We have fitted pulsation models, including magnetic perturbation effects and
assuming dipole magnetic geometries, to the frequency spectrum of the rich multi-periodic 
 star HD~134214, which pulsates with the shortest periods known among the roAp stars. 
We have found that a better fitting is obtained when we fit 8 frequencies common to 
the MOST photometry \citep{gru11} and radial velocity data obtained by \citep{ku07}, 
rather than fitting the 10 frequencies obtained by MOST.
This might indicate that the two new frequencies obtained by \citet{gru11} have 
high surface degrees of $\ell > 3$.

Among the best fitting models found  in our grid for various input physics, we found 
that a model with a mass of $1.65\,{\rm M}_\odot$,  effective temperature $\log T_{\rm eff} = 
3.8584$, luminosity $\log L/{\rm L}_\odot = 0.9250$, and polar magnetic field strength 
$B_{\rm p} = 4.1$\,kG is the best match to the data.  The mean deviation of the fit is 
$1.23\,\mu$Hz, which is slightly larger than the expected numerical errors for 
the model frequencies.  
Non-dipole components in the magnetic field of HD~134214 may be additional error sources.

The effective temperature and luminosity of the best model are consistent with those 
derived from multicolour photometry and the Hipparcos parallax. We note that the 
parameters are nearly the same as those of the best model for HD~24712 obtained by 
\citet{sa10}.  

The principal frequency, with a photometric amplitude ten times larger than that 
of the other frequencies, is identified as a deformed dipole mode. 
The five frequencies with the next largest amplitudes are identified 
as $\ell_{\rm m}=2$ modes. 
In our model, the radial velocity variations of these modes have a 
quasi-node around $\log \tau \approx -3$, where oscillation phase shifts by $\sim 2$ 
radians. The radial velocity amplitude increases rapidly above the node.
These properties qualitatively agree with the data from radial velocity 
measurements of H$\alpha$ bisector and lines of rare earth elements.

Although the model frequencies and amplitude distribution in the atmosphere are 
consistent with observations of HD~134214, the excitation mechanism is still unknown.
The kappa-mechanism in the hydrogen ionization 
zone, e.g., appears to be too weak to excite oscillations in the observed 
frequency range, above the acoustic cut-off frequency.  

Finally, we note that Gruberbauer (2011 in preparation) has recently 
developed a new probabilistic method for finding a best model in reproducing 
observed frequencies.
One of the advantages of the method is its ability of taking into account  the 
possibility of the presence of systematic errors, which we disregarded completely 
in this paper.
Applying such a probabilistic method to various roAp stars including HD~134214 
itself might reveal systematic errors in the model frequencies as a function of 
the frequency and/or magnetic field strength.

\section*{Acknowledgments}
We are grateful to Don Kurtz, the referee of this paper, 
for his helpful comments and suggestions.
MG has received financial support from an NSERC Vanier scholarship.
WWW was supported by the Austrian Research Fond (P22691-N16).

\bsp

\label{lastpage}


\begin{thebibliography}{}
\bibitem[\protect\citeauthoryear{Adelman}{2000}]{ad00}
Adelman S. J., 2000, A\&AS, 146, 13

\bibitem[\protect\citeauthoryear{Alexander \& Ferguson}{1994}]{alex94}
Alexander D. R., Ferguson J. W., 1994, ApJ, 437, 879

\bibitem[\protect\citeauthoryear{Audard et al}{1998}]{aud98}
Audard N., Kupka F., Morel P., Provost J., Weiss W. W., 
1998, A\&A,  335, 954

\bibitem[\protect\citeauthoryear{Balmforth et al.}{2001}]{bal01}
Balmforth N. J., Cunha M. S., Dolez N., Gough D. O., Vauclair S.,
2001, MNRAS, 323, 362

\bibitem[\protect\citeauthoryear{Balona et al.}{2011a}]{bal11a}
Balona L. A., Cunha M. S., Kurtz D. W., Brand\~ao I. M., Gruberbauer M.,
 Saio H., \"Ostensen R., Elkin V. G., Borucki W. J., Christensen-Dalsgaard J., 
Kjeldsen H., Koch D. G., Bryson S. T., 2011a, MNRAS,  410, 517

\bibitem[\protect\citeauthoryear{Balona et al.}{2011b}]{bal11b}
Balona L. A.,  Cunha M. S., Gruberbauer M.,  Kurtz D. W.,
Saio H.,  White T. R.,  Christensen-Dalsgaard J., Kjeldsen H.,
Christiansen J. L.,  Hall J. R.,  Seader S. E., 2011b, MNRAS, in press

\bibitem[\protect\citeauthoryear{Bigot \& Kurtz}{2011}]{bk11}
 Bigot L., Kurtz D. W., 2011, A\&A, in press (arXiv:1110.0988)

\bibitem[\protect\citeauthoryear{Bigot \& Dziembowski}{2002}]{bd02}
 Bigot L., Dziembowski W.A., 2002, A\&A, 391, 235


\bibitem[\protect\citeauthoryear{Bigot et al.}{2000}]{big00}
 Bigot L., Provost J., Berthomieu G., Dziembowski W. A., Goode P. R.,
 2000, A\&A, 356, 218

\bibitem[\protect\citeauthoryear{Cameron et al.}{2006}]{ca06}
Cameron C., Matthews J. M.,  Rowe J. F.,  Kuschnig R.,  Guenther D. B.,
 Moffat A. F. J.,  Rucinski S. M.,  Sasselov D.,  Walker G. A. H.,  Weiss W. W.,
 2006, Comm. Asteroseismology,  148,  57

\bibitem[\protect\citeauthoryear{Chandrasekhar}{1961}]{ch61}
Chandrasekhar S., 1961, Hydrodynamic and Hydromagnetic Stability (Oxford: Clarendon) 

\bibitem[\protect\citeauthoryear{Cox}{1984}]{cox84}
 Cox J. P., 1984, ApJ, 280, 220

\bibitem[\protect\citeauthoryear{Cunha}{2002}]{cu02}
 Cunha M. S. 2002, MNRAS, 333, 47

\bibitem[\protect\citeauthoryear{Cunha}{2006}]{cu06}
 Cunha M. S. 2006, MNRAS, 365, 153

\bibitem[\protect\citeauthoryear{Cunha \& Gough}{2000}]{cg00}
Cunha M. S., Gough D. O., 2000, MNRAS, 319, 1020

\bibitem[\protect\citeauthoryear{Dziembowski \& Goode}{1996}]{dg96}
 Dziembowski W. A., Goode P. R., 1996, ApJ, 458, 338

\bibitem[\protect\citeauthoryear{Gautschy et al.}{1998}]{gsh98}
 Gautschy A., Saio H., Harzenmoser H., 1998, MNRAS, 301, 31

\bibitem[\protect\citeauthoryear{Gruberbauer et al.}{2011}]{gru11}
Gruberbauer M., Huber D., Kuschnig R., Weiss W. W., Guenther D., 
Matthews J. M., Moffat A. F. J., Rowe, J. F., Rucinski S. M., Sasselov D., 
Fischer M., 2011, A\&A in press

\bibitem[\protect\citeauthoryear{Gruberbauer et al.}{2008}]{gru08}
Gruberbauer M., Saio H., Huber D., Kallinger T., Weiss W. W., Guenther D. B.,
Kuschnig R., Matthews J. M., Moffat A. F. J., Rucinski S., Sasselov D., 
Walker G. A. H., 2008, A\&A, 480, 223

\bibitem[\protect\citeauthoryear{Henyey, Vardya \& Bodenheimer}
{Henyey et al.}{1965}]{heny65}
Henyey L., Vardya M.S., Bodenheimer, P., 1965, 142, 841

\bibitem[\protect\citeauthoryear{Huber et al.}{2008}]{hub08}
Huber D., Saio H., Gruberbauer M., Weiss W. W., Rowe J. F., Hareter, M., 
Kallinger T., Reegen P., Matthews J. M., Kuschnig R., Guenther D. B., 
Moffat A. F. J., Rucinski S., Sasselov D., Walker G. A. H., 2008, A\&A,
483, 239

\bibitem[\protect\citeauthoryear{Hubrig, North \& Mathys}{Hubrig et al.}{2000}]
{hubrig00}
Hubrig S., North P., Mathys G., 2000, ApJ, 539, 352

\bibitem[\protect\citeauthoryear{Iglesias \& Rogers}{1996}]{opal95}
 Iglesias C. A., Rogers R. J., 1996, ApJ, 464, 943

\bibitem[\protect\citeauthoryear{Kochukhov \& Bagnulo}{2006}]{koc06}
Kochukhov O., Bagnulo S., 2006, A\&A, 450, 763

\bibitem[\protect\citeauthoryear{Kreidl}{1985}]{kreidl85}
Kreidl T. J., 1985, IBVS, 2739, 1

\bibitem[\protect\citeauthoryear{Kreidl \& Kurtz}{1986}]{kk86}
Kreidl T. J., Kurtz D. W., 1986, MNRAS, 220, 313

\bibitem[\protect\citeauthoryear{Kreidl et al.}{1994}]{kr94}
Kreidl T. J., Kurtz D.W., Schneider H., vanWyk F., Roberts G., Marang F.,
Birch P. V., 1994, MNRAS, 270, 115

\bibitem[\protect\citeauthoryear{Kurtz}{1982}]{ku82}
 Kurtz D. W., 1982, MNRAS, 200, 807

\bibitem[\protect\citeauthoryear{Kurtz et al.}{2005}]{ku05}
 Kurtz D. W., Cameron C., Cunha M. S., Dolez N., Vauclair G., Pallier E.,
 Ulla A., Kepler S. O., da Costa A., Kanaan A. {\it et al.}, 2005, MNRAS, 
 358, 651

\bibitem[\protect\citeauthoryear{Kurtz et al.}{2011}]{ku11}
Kurtz D. W., Cunha M. S., Saio H., Bigot L., Balona L. A., Elkin V. G., Shibahashi H., 
Brandao I. M., Uytterhoeven  K., Frandsen S., Frimann S., Hatzes  A., Lueftinger T., 
Gruberbauer  M., Kjeldsen H., Christensen-Dalsgaard J., Kawaler S. D., 
2011, MNRAS, 414, 2550

\bibitem[\protect\citeauthoryear{Kurtz et al.}{2007}]{ku07}
Kurtz D. W., Elkin V. G.,  Mathys G., van Wyk F., 
 2007, MNRAS, 381, 1301

\bibitem[\protect\citeauthoryear{Kurtz et al.}{2006a}]{ku06}
Kurtz D. W., Elkin V. G., Cunha M. S., Mathys G.,  Hubrig S.,  Wolff B., 
Savanov I., 2006a, MNRAS, 372, 286

\bibitem[\protect\citeauthoryear{Kurtz, Elkin \& Mathys}{Kurtz et al.}{2006b}]{ku06b}
Kurtz D. W., Elkin V. G., Mathys G.,  2006b, MNRAS, 370, 1274

\bibitem[\protect\citeauthoryear{Mathys et al.}{1997}]{mat97}
Mathys G., Hubrig S., Landstreet J.D., Lanz T., Manfroid J., 1997,
A\&AS, 123, 353

\bibitem[\protect\citeauthoryear{Matthews, Kurtz \& Wehlau}{Matthews et al.}{1987}]
{mat87}
Matthews J. M., Kurtz D. W., Wehlau W. H., 1987, ApJ, 313, 782

\bibitem[\protect\citeauthoryear{Mkrtichian \& Hatzes}{2005}]{mkr05}
Mkrtichian D. E., Hatzes A. P., 2005, A\&A, 430, 263

\bibitem[\protect\citeauthoryear{Mkrtichian et al.}{2008}]{mkr08}
Mkrtichian D. E., Hatzes A. P., Saio H., Shobbrook R. R., 2008, A\&A, 490, 1109

\bibitem[\protect\citeauthoryear{Renson \& Manfroid}{2009}]{ren09}
Renson P., Manfroid J., 2009, A\&A, 498, 961


\bibitem[\protect\citeauthoryear{Roberts \& Soward}{1983}]{rs83}
Roberts P. H., Soward A.,  1983,   MNRAS, 205, 1171

\bibitem[\protect\citeauthoryear{Ryabchikova, Kochukhov \& Bagnulo}
{Ryabchikova et al.}{2008}]{rya08}
Ryabchikova T., Kochukhov O., Bagnulo S., 2008, A\&A, 480, 811

\bibitem[\protect\citeauthoryear{Ryabchikova et al.}{2007}]{rya07}
Ryabchikova T., Sachkov M., Kochukhov O., Lyashko D., 2007, A\&A, 473, 907


\bibitem[\protect\citeauthoryear{Saio}{2005}]{sa05}
Saio H., 2005, MNRAS, 360, 1022

\bibitem[\protect\citeauthoryear{Saio}{2008}]{sa08}
Saio H., 2008, JPhCS, 118, 012018

\bibitem[\protect\citeauthoryear{Saio \& Gautschy}{2004}]{sg04}
Saio H., Gautschy A., 2004, MNRAS, 350, 485

\bibitem[\protect\citeauthoryear{Saio, Ryabchikova \& Sachkov}{Saio et al.}{2010}]{sa10}
Saio H., Ryabchikova T., Sachkov M.,  2010, MNRAS, 403, 1729

\bibitem[\protect\citeauthoryear{Shibahashi}{1983}]{shb83}
Shibahashi H., 1983, ApJ, 275, L5

\bibitem[\protect\citeauthoryear{Shibahashi \& Saio}{1985}]{ss85}
Shibahashi H., Saio H., 1985, PASJ, 37, 245

\bibitem[\protect\citeauthoryear{Sousa \& Cunha}{2011}]{sc11}
Sousa J. C., Cunha M. S. 2011, MNRAS, 414, 2576

\bibitem[\protect\citeauthoryear{Unno \& Spiegel}{1966}]{un66}
Unno W., Spiegel E. A., 1966, PASJ, 18, 85
\end{thebibliography}
\end{document}